\documentclass [dvips, 12pt] {article} 
\usepackage {graphicx}
\usepackage {pifont}
\newcommand {\RM} {\ensuremath {\rho (770)} }

\newcommand {\pion} {\ensuremath {{\pi^0}} }
\newcommand {\kaon} {\ensuremath {{K^0}} }
\newcommand {\ip} {\ensuremath {{\pi^\pm}} }
\newcommand {\ak} {\ensuremath {{K^\pm}} }

\newcommand*{\Approx}[1]{\ensuremath {\approx #1}}

\newcommand {\cu} {\normalsize \textcircled {\em \footnotesize u}}
\newcommand {\cs} {\normalsize \textcircled {\em \footnotesize s}}

\newcommand {\cub} {\normalsize \textcircled {\em \footnotesize $\bar{u}$}}
\newcommand {\csb} {\normalsize \textcircled {\em \footnotesize $\bar{s}$}}

\begin {document}

\begin {center}
{\Large The six-quark structure of long-lived $D$-mesons and $\psi/J$ \\ }
     \par\bigskip 
{\large Oleg~A.~Teplov}\par\smallskip 
Institute of metallurgy and materials science 
of the Russian Academy of Sciences, 
Moscow.
\par e-mail: teplov@ultra.imet.ac.ru 
\end {center}

\begin{abstract}

The harmonic quarks are applied to the analysis 
of quark structures of long-lived $D$-mesons and $\psi/J$. 
In the harmonic quark model the quarks of 
long-lived mesons ($\pi^0$, $\pi^\pm$, $K^\pm$, $K^0$, $B^0$, $B^\pm$) 
are weakly relativistic objects with the excess energy 
of valent quarks less than 7 MeV. In contrast to this, the long-lived $D$-mesons 
in two-quark standard have the apparent excess energies of a few hundred MeV.
It was established that all long-lived $D$-mesons and $\psi/J$ have 
the six-quark structures.
There are two valent quarks and two additional neutral 
quark-antiquark $u$-pairs. 
The quark structures of $D^0$, $D^\pm$, $D_s^\pm$ and $\psi/J$ are given. 
The excess energies of six-quark $D$-mesons also does not exceed 7 MeV. 
A six-color quark composition and the absence of two-quark $D$-mesons
in the context of QCD are discussed.

\end{abstract}

\section {Introduction}

\quad The harmonic quarks have proved that they are the powerful 
and effective tool
for investigation of energy structure of hadrons~\cite{my1}--\cite{my5}.
This tool works equally well 
with both masses of long-lived hadrons and resonances. 
The hypothesis, that there is a simple analytical association of quark masses, 
was not created out of thin air, but it is actually a conclusion 
of observable regularity 
of a long-lived mesons mass spectrum~\cite{my1}.
The harmonic quarks/oscillators allowed to decrypt the spectrum of meson
masses with an open charm~\cite{my4}, and to interpret~\cite{my5} the precise 
data of collaborations CLEO and BELLE for mass differences 
of the charmed mesons~\cite{cleo, belle}.
Thus, it may be considered proven that these differences are strictly quantized
by harmonic quarks rest masses.

In the present work, 
we shall investigate the energy states and quark structures 
of the long-lived charmed mesons.
 This problem becomes important when we try to interpret 
the structures of long-living charmed mesons using the harmonic quarks.  
As was mentioned earlier in~\cite{my4}, the mass of long-lived $D$-mesons 
($D^0$, $D^\pm$ and $D_s^\pm$) have an excess mass more than total mass 
of two valent harmonic quarks. 
At the same time, the harmonic quark concept 
and especially the calculation of their masses~\cite{my1}
is set the fact that the quarks in long-lived mesons, 
such as \pion, \ip, \ak and $b^\pm$, are weakly relativistic objects.
The excess  energy of these mesons, i.e. energy  above rest mass 
of two valent quarks, do not exceed 7 MeV~\cite{my1} 
and  can be easily defined\footnote{The masses 
of the harmonic quarks $d$, $u$,
 $s$, $c$ and $b$ are correspondingly next~\cite{my2, my3}: 
 28.811, 105.441, 385.89, 1412.28 and 5168.7. 
The energy of completed $u$-oscillator $\cu\cub$\, is equal 134.25 MeV. 
The notation is as in~\cite{my2, my3, my4}}.
{
\begin{center}
Table 1. The quark structures of long-lived mesons and the rest masses of their quarks.  

  \medskip\small 
  \begin{tabular}{|c|c|c|c|c|}
    \hline
  & Quark structure & Mass & Mass of quarks & Excess energy, \\
 Meson       &  of meson & of meson & or oscillator, & MeV  \\
       &  & &  MeV/$c^2$  &  \\
\hline
   \pion & \cu\cub\, & 134.98 & 134.25 & 0.73    \\
\hline
   $\pi^+$ & $u\bar{d}$ & 139.57 &134.25 & 5.32    \\
\hline
   $K^+$ & $u\bar{s}$ & 493.66 & 491.33 & 2.33  \\
\hline
   $D^0$ & $c\bar{u}$ & 1864.5 & 1517.72 & 346.78  \\
\hline
   $D^+$ & $c\bar{d}$ & 1869.3 & 1517.72 & 351.58 \\
\hline
   $D_s^+$ & $c\bar{s}$ & 1968.2 & 1798.17 & 170.03    \\
\hline
   $\psi/J$ & $c\bar{c}$ & 3096.916 & 2824.56 & 272.35    \\ 
\hline

  \end{tabular}

\end{center}
}
\par \bigskip

As against them in long-lived $D$-mesons ($D^0$, $D^\pm$ and $D_s^\pm$), 
an excess energy 
differs approximately per two order and achieves values \Approx 350 MeV (see tab.1). 
Such values of excess energies are characteristic for resonances,
but not for ground meson states which decay 
because of a weak interaction.
There are no visible reasons, which hinder a decrease of this
 "superfluous" energy and a pion emission in the result of strong interaction.

\begin {figure} [htb]
\begin {center}
\includegraphics [scale =.7] {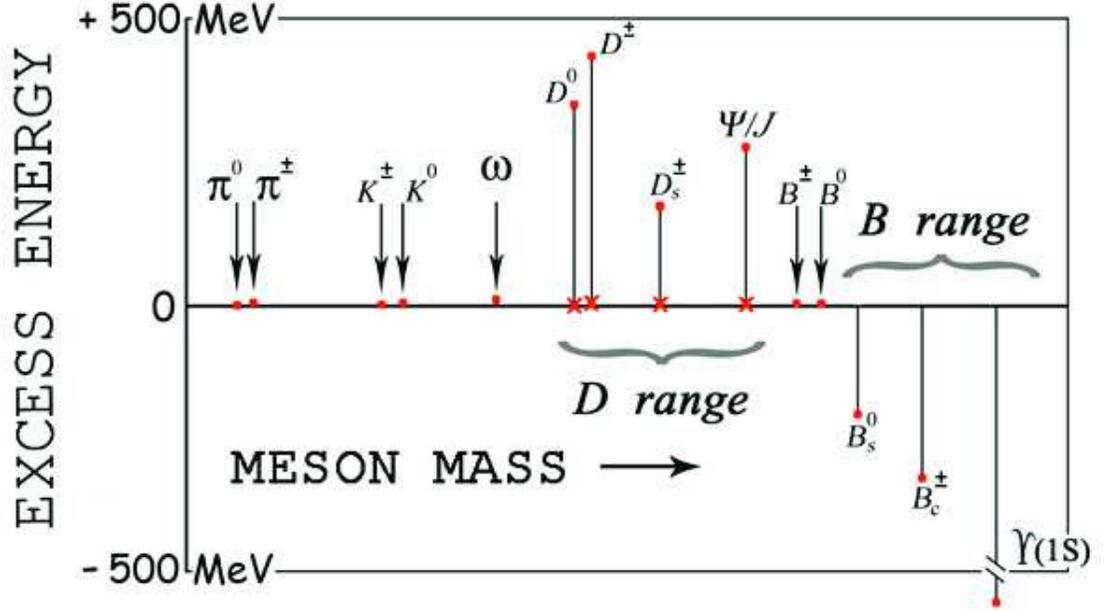}
\par
\caption [] {The excess energy of long-lived and first vector mesons.

\label {fig:excess}}
\end {center}
\end {figure}

The fig.\ref{fig:excess} is shown a scheme, which explains 
this problem. It demonstrates excess energy for long-lived mesons 
(\pion, \ip, \ak, \kaon, $D^0$, $D^\pm$, $D_s^\pm$, $B^\pm$, 
$B^0$, $B_s$, $B_c^\pm$) 
and the first vector mesons with hidden flavors 
($\omega$, $\psi/J$, $\Upsilon$(1S)).

\par
\par

Here it is necessary to explain, 
that we understand as excess energy (EE). 
The excess energy for \ip, \ak, $D^0$, $D_s^\pm$, $B^\pm$, $B_s$, $B_c^\pm$ is 
a difference between a mass of meson and the summary mass 
of the valent quarks. 
Therefore, for example, for \ip and  $D_s^\pm$ it is
\medskip
 
EE($\pi^\pm$) = {$M_{\pi^\pm}$} - {$M_u$} - {$M_d$}  
 ~~~~~~~~and  ~~~~~~~EE($D_s^\pm$) = $M_{D_s^\pm}$ - $M_c$ - $M_s$.  
\medskip
  
For the second states of hadronic doublets to these values 
are added a mass differences of proper doublets.
The choice of this scheme for second states has certain reasons
  that discussed in~\cite{my2, my4}.    
So, for example, the excess energy for \pion and $D^\pm$ corresponds 
to the following equalities:
\medskip

EE($\pion$) = EE($\pi^\pm$) - ($M_{\pi^\pm}$  - $M_{\pion}$) 
~~~~and~~~~  EE($D^\pm$) = EE($D^0$) -  ($M_{D^\pm}$  - $M_{D^0}$).
\medskip

For the vector mesons, in which the standard model guesses 
a presence of quark-antiquark pair of corresponding flavor, 
the excess energies are equal:
\medskip

 EE($\omega$) = $M_{\omega}$ - 2$M_s$;
 ~~~~~~EE($\psi/J$) = $M_{\psi/J}$ - 2$M_c$;
 ~~~~~~EE($\Upsilon$(1S) = $M_{\Upsilon(1S)}$ - 2$M_b$.
\medskip

On the fig.\ref{fig:excess}, we may see three groups of states for long-lived mesons.
In the first group an excess energy is small and plus, 
i.e. quarks in this group of mesons are weak relativistic objects. 
In the second group a large negative excess energies 
are observed for the mesons with $b$ quarks. 
The binding energy in these mesons is great and it defines 
by excess energy itself of $b$-quark, i.e. its donor properties~\cite{my4}. 
In the given work, the object of investigation is third group,
which has the mesons with $c$ quarks and
with great positive excess energies. 

  At sequential using of the harmonic quark model, we should suppose 
  that the quarks in all long-lived mesons are weakly relativistic objects. 
From here follows, that {\bf the greatest part of excess energy in mesons
$D^0$, $D^\pm$ and $D_s^\pm$ should be concentrated in masses of quarks 
and the complete oscillators}, i.e. a quark structure of these mesons 
should be more complex, than in the standard quark model is guessed. 
  
Let us note also that hereinafter we shall use only experimental masses
 of mesons~\cite{part}.

\section {The quark structures of long-lived charmed mesons}

\quad As it was noted earlier for pions~\cite{my1, my2}, electromagnetic split 
of their masses is bound with the valent $d$-quark,
as at approximately equal momentum of quarks in pions 
the lightest harmonic $d$-quark should have the greater kinetic energy,
than a $u$-quark. This standing is valid and for various kaons,
baryons and for $D$-mesons also. {\em There are no the meson multiplets
with the inverse order of mass split}. 
From a position of the harmonic levels 
as a gang of harmonic oscillators,
mesons $D^0$ and $D^\pm$ belong to one level~\cite{my4}, 
but their second valent quarks
accordingly $u$ and $d$ have the rest masses 
distinguishing approximately
in 3,6 times.
Both mesons are in one potential well, but in view 
of the previously mentioned, the mass of a meson  $D^\pm$ with the valent $d$-quark 
should be more masses of a meson  $D^0$. Really, the difference 
of their masses is same as and at pions.

Now we shall estimate the excess energy of quarks in a meson  $D^0$, 
using their values for the first long-lived mesons of other flavors 
(\pion, \ak and $b^\pm$). 
These values accordingly the following: 0.725; 2.33 and 4.89 MeV.
The choice of these mesons, instead of three other long-lived mesons 
(\ip, \kaon and $b^0$), is determined by that the second group 
has the valent $d$-quark and is more suitable for definition 
of excess energy in a meson $D^\pm$.
We simply do not have 
other variants for $D^0$. Taking into account the electrical charges 
of mesons and heavy quarks in mesons ($D^0$ and $b^\pm$), 
we make a proportion:

\begin{equation}\label{EED0}
EE(D^0)/EE(\pion) = EE(b^\pm)/EE(K^\pm)   
\end{equation}

The ratio (\ref{EED0}) is balanced enough. Both in left-hand, 
and in the right ratios are a mesons with massive quarks 
through generation, i.e. quarks have same charges 
($b$- and $s$-quarks for the right ratio).
From here, we find EE($D^0$) = 1.52 MeV.
Then the sum of rest masses of quarks $D^0$ 
is approximately equal 1863.0 MeV.
The energy above mass of the valent quarks ($c$ and $\bar{u}$)
shall be equal approximately 345.3 MeV.
This energy should be a rest mass of electrically neutral 
and colorless group of quarks.
It can consist only from $u$- and $d$-quarks
and their oscillators as the energy of {$s$-oscillator} more and 
is equal 491.33 MeV.

Really, the analysis has shown, that this energy 
is matched with equivalent mass of the most simple symmetric quark
 combination (minimum of quarks and oscillators): 

$u\bar{u}$ + \cu\cub

The energy of last group is equal 345.13 MeV.
Next combination with such energy  contains already 
eight quarks, including six $d$-quarks.
This group should be rejected for many reasons.
Here is one of them. 
At least four $d$-quark are bound with each other, 
i.e. they are the free, 
and each of them demands additional energy approximately same
as the valent $d$-quark surveyed above. 
This energy is more on one order 
than our estimation of excess energy in $D^0$. 
The next rejected combination have already 12 $d$-quarks.
That is all, there are not one more other neutral quark group.
Thus, the first combination with the minimum number of quarks 
is single acceptable neutral group. 
From here we define, 
that excess energy of quarks in $D^0$ is equal 1.65 MeV 
and the mass of all quarks is equal 1862.85 MeV.
At present, we may only carefully congratulate itself with this result. 
The quark composition of the $D^0$ is next:

\begin{equation}\label{D0}
c\bar{u} + u\bar{u} + \cu\cub\,   
\end{equation}

In~\cite{my2} we noted that  completely neutral shells 
(colorless electrically neutral shells) from six quarks
of one flavor correspond to a QCD and consequently
should be inconvertible  configurations, 
as well as integrating of quarks in completely neutral pairs.
Really, we managed to show, that shells from six quarks 
are present at some resonances~\cite{my2, my3}. 
However, inconvertible structure from six quarks in long-lived meson
is the unexpected result which necessarily in the further
to discuss and to check. 
 
The replacement of \cu\cub-oscillator in $D^0$ on quarks $u\bar{d}$ 
with the same mass give to us at once the quark structure of $D^+$
with excess  energy 6.45 MeV:

\begin{equation}\label{Dpm}
c\bar{d} + u\bar{u} + u\bar{u}   
\end{equation}

The EE in $D^\pm$ is little more, than EE 
with the $d$ valent quarks in \ip and \kaon 
(5.32 and 6.34 MeV accordingly).
It is possibly bound simply with larger number of quarks in $D^\pm$.

Now we shall estimate the excess energy in a meson $D_s^\pm$. 
On the one hand a EE of $D_s^\pm$ should be more, 
than for $D^0$ (1.65 MeV) because the valent quarks 
of the $D_s^\pm$ are unipolar, but on the other hand it should be less,
than for meson $D^\pm$ (6.45 MeV) because $s$-quark on one order is heavier
than $d$-quark. The \ak with EE 2.33 MeV 
can be by the analog for $D_s^\pm$, however $D_s^\pm$, 
probably, has more quarks. 
The reader, we shall take mean quantity on three values since 
it is all that we can. Therefore, mean value is equal 3.48$^{+3}_{-1.8}$ MeV
and the rest mass of quarks is equal about 1964.72 MeV.
The limitations (+3 and -1.8) follow from EE($D^\pm$) and EE($D^0$).
The surplus of mass above the valent quarks is equal 166.6 MeV.
Simple exhaustive search of variants has shown that energy 166.6
in boundaries $\pm$3.0 MeV is not interpreted as neutral group 
from $u$- and $d$-quarks and their oscillators. 

However, we have one more chance.
In~\cite{my4} it was noted, that in the strong central field
of $c$-quark the mass rank $s$- and $u$-quarks 
can be changed down to their state
in complete harmonic oscillators.
For $s$-quark, this state 
is \cs ~with energy 245.666 MeV. 
The surplus of mass above 
the valent quarks in this case will be equal 306.8 MeV.
This energy can be precisely represented also by four $u$-quarks:

\begin{equation}\label{EEDs}
for ~D_s^{+}~~~~\bar{u} + \cu\, + \cu\cub ~~~~~~~for ~D_s^{-}~~~~u + \cub\, + \cu\cub  
\end{equation}

The rest mass of these quarks is 306.82 MeV.
Certainly, we was lucky with accuracy, but nevertheless 
the quark structure of a meson $D_s^+$ may be next:

\begin{equation}\label{Ds}
c\csb\, + \cu\, + \bar{u} + \cu\cub\,   
\end{equation} 

From (\ref{Ds}) follows that excess energy of quarks in $D_s^\pm$
is equal 3.44 MeV, and the rest mass of quarks is equal 1964.76 MeV.

Thus, we have received compositions 
with six harmonic quarks for all long-lived $D$-mesons of base levels.
The rest mass of all quarks for these mesons are spotted 
with precision about 0.1 MeV (see tab.2). Therefore, the error of definition 
of EE will about 0.4 MeV, because the experimental mass error
is equal 0.4 MeV~\cite{part} and it gives the greatest contribution.

\subsection {\bf Application of harmonic quarks to $\psi/J$}

\quad In spectrum of charmonium, the $\psi/J$ 
is first long-lived meson with hidden charm ($\eta_c$ is a resonance 
with width about 25 MeV). 
Therefore, with above-said point of view, we may expect 
that its quarks are also the weak relativistic objects. 
The mass difference between $\psi/J$ and rest mass of two $c$-quarks 
is equal 272.35 MeV and may consist only of $u$- and $d$-quarks.
The excess energy of $\psi/J$ may be estimated simply as mean value
between the excess energies (now known to us) 
of $D^0$, $D^\pm$ and $D_s^\pm$,
i.e. as 3.85 MeV (see tab.2). 
From here, the rest mass of neutral quark group is equal 268.50 MeV.
We was again lucky with accuracy. This mass is simply energy 
of two $u$-oscillators (268.50 MeV).
Therefore, the quark structure of $\psi/J$ may be noted down 
as symmetric group out of six quarks:

\begin{equation}\label{psi}
c\bar{c} + \cu\cub\, + \cu\cub\,  
\end{equation}

As it was mentioned above, a $c$-quark has acceptor properties 
and each heavy quark can capture one of $u$-oscillators. 
Then the configuration (\ref{psi}) may be represented as the molecule 
of two quark atoms:

\begin{equation}\label{molecula}
c + \cu\cub\,  ~~and  ~~\bar{c} + \cu\cub\,   
\end{equation}

The EE of $\psi/J$ is equal 3.85 $\pm$ 0.15 MeV.
Indicated error  corresponds with precision which the quark masses
 is defined~\cite{my3}.
Summarized results are given in table 2 and the new positions of 
excess energy for $D$-mesons are pointed out  on fig.1 by crosses.

\begin{center}
Table 2. The quark structure of long-lived mesons, the rest mass
and excess energies of their quarks.  

  \medskip\small 
  \begin{tabular}{|c|c|c|c|}
    \hline
Long-lived & Quark structure & Rest mass & Excess energy, \\
mesons       &  of meson & of quarks, MeV/$c^2$  & MeV  \\
    \hline
   $\pi^+$ & $u\bar{d}$ & 134.25$\pm$0.007 & 5.32$\pm$0.007 \\
\hline
   $K^+$ & $u\bar{s}$ & 491.33$\pm$0.025 & 2.33$\pm$0.030  \\
\hline
   $D^0$ & $c\bar{u}$ + \cu\cub\, + $u\bar{u}$ & 1862.85$\pm$0.09 
   & 1.65$\pm$0.41  \\
\hline
   $D^+$ & $c\bar{d}$ + $u\bar{u}$ + $u\bar{u}$ & 1862.85$\pm$0.09 
   & 6.45$\pm$0.41 \\
\hline
   $D_s^+$ & $c$\csb\, + \cu\, + $\bar{u}$ + \cub\cu\, & 1964.76$\pm$0.10
   & 3.44$\pm$0.51 \\
\hline
   $\psi/J$ & $c\bar{c}$ + \cu\cub\, + \cu\cub\, & 3093.06$\pm$0.15 
   & 3.85$\pm$0.15    \\ 
\hline
    $b^+$ & ${u\bar{b}}$  & 5274.11$\pm$0.26 & 4.89$\pm$0.56 \\
\hline

  \end{tabular}

\end{center}

\par \bigskip

The errors of excess energy in tab.2 are defined 
as squared errors between experimental mass errors 
of mesons~\cite{part} and the errors of rest mass. 

\section {Discussion}

\quad This and earlier done investigations 
of $D$-meson structures with the using 
of the harmonic quarks leads us to the conclusion about discovery  
of new unknown phenomenons in a microcosm.
At first, we discovered that the some $D$-transitions 
are precisely quantized by the masses of harmonic quarks~\cite{my5}.
Secondly in present work, we found the unusual quark structures 
of long-lived ground $D$-mesons and $\psi/J$.

What are the probability that these discoveries are true?
This matter was investigated earlier in~\cite{my3, my4}.
 So, the probability of an accidental coincidence of actual 
and model spectrums 
of hadrons up to 1000 MeV is estimated in~\cite{my3} 
as value less than $10^{-6}$. 
For actual and model spectrums of mesons with open charm~\cite{my4}, 
the variance of correspondence is six times less
 in comparison with~\cite{my3}.
Here already the probability of an accidental coincidence
is approximately equal $10^{-12}$.
The independent value from two works is less than $10^{-18}$.
After~\cite{my5} and present article, it becomes progressively less.

\subsection {Composition Six}

\quad We have detected the six-quark compositions in long-lived $D$-mesons
and $\psi/J$. 
The author supposes that all they are the six-color quark
compositions with full set of colors (red, green and blue) and anticolors.
The compositions of $D^\pm$ and $\psi/J$ with two identical 
quark-antiquak pairs
testifies in favour of this assumption.
At ground state these pairs must have the same quantum numbers (L=0, S=0) 
except color only.

\begin {figure} [htb]
\begin {center}
\includegraphics [scale =.5] {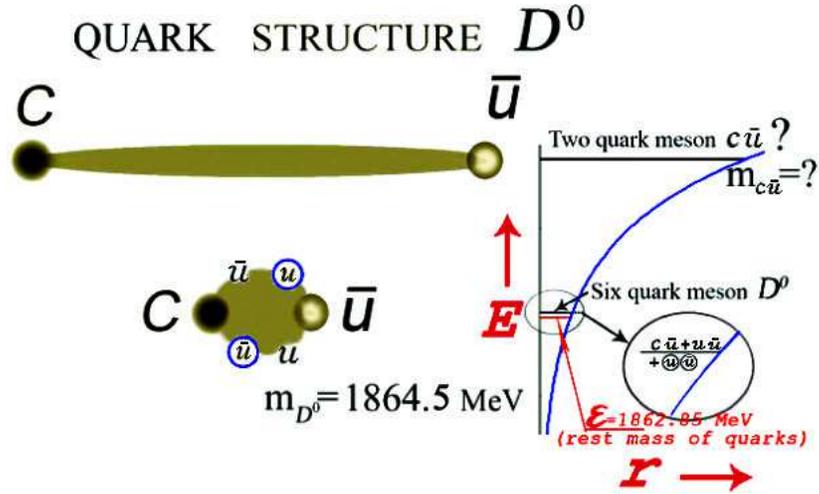}
\par
\caption [] {The illustrative scheme of strong interaction in $D^0$ 
for two- and six-quark configurations.
\label {fig:d0}}
\end {center}
\end {figure}

The scheme on fig.\ref{fig:d0} illustrates two type of quark structure 
for $D^0$: 

1. two-quark strong coupling with long tube;

2. six-quark strong interaction with short tubes.

\par
\par

The six-color composition can explains a too large excess energy
in {$D$-meson} and removes the contradiction between two groups 
of long-living mesons (see fig.\ref{fig:excess}).
The contradiction is inevitable only for two quark modes
in standard quark model. Two-quark long-living meson must often be in state
with very extended color tube and maximum potential energy.
There are many chance for condensation of EE in the form 
of quark-antiquark pairs on acceptor trap of $c$-quark. 
It is one of possible ways  the six-quark composition might be formed.

At the same time, perhaps, the two-quark $D$-mesons not exist in the nature.

\subsection {QCD and $D$-mesons}
Is this result unexpected? 
In addition, how nevertheless all this is according with QCD?

The heavy $c$-quark introduces an asymmetry in $D$-mesons 
and automatically becomes the valent quark together 
with its anticolor partner.
Any other quark-antiquark pair will consist from $u$-quarks 
(see tab.2) with probably another color and an anticolor, which differ 
from colors of $c$-quark and its partner. A presence in a meson 
of others completely neutral quark-antiquark pairs remains hidden
for standard quark model, as well as, for example, 
a sea quark-antiquark pairs. However, for $D$-mesons these pairs
are functional, since they are necessary units 
of the complete color group, i.e. the composition 6.
The shells with configuration 6 in the harmonic quark model
were expected and detected earlier in~\cite{my3}. 
These shells are guessed in structures \RM, $K^*$, $\varphi$, $a$(1474) 
and $p\bar{p}$.
This property is not unique for the charmed mesons.
The distinctive feature of $D$-spectrum from light and strange mesons is 
{\bf \em the full absence of long-living $D$-mesons with one quark-antiquark 
pair}.

Two-quark long-lived $D$-meson in the harmonic model should have
the mass about 1520 MeV with excess energy in range 1-7 MeV
similar mesons \ip, \ak, $B^\pm$.
However the charmed mesons with such masses are not observed, 
we have $D$-mesons with masses about 1870 MeV and, 
as shown in this work, in six-quark configurations 
with the mentioned above excess energies 
(1.65 for $D^0$ and 6.45 MeV for $D^\pm$).

{\large \em The next scenario is most probable}.
 
The QCD and absence of two-quark $D$-mesons tell to us, 
that color interaction of a $c$-quark and its partner is 
insufficient for formation of long-lived two-quark meson.
It means, an exchange of color gluons between $c$- and $u$-quarks 
is too small. 
A weak gluon activity can give a bad allocation and bad compensation 
of colors on space-time and a break of asymptotic freedom.
Then, as the result,there will not of an enough strong coupling 
between quarks for formation of two-quark $D$-mesons. 
Other quarks have two-quark long-living mesons
(except $t$-quark) and,
hence, enough intensive gluon exchange for an establishing 
of antishielding and colorless superposition 
$R\bar{R}$ + $G\bar{G}$ + $B\bar{B}$,
i.e. a formation of strong coupling. 

Just properties of $c$-quark show up in a weakening of gluon exchange. 

Though in QCD an emission rate of gluons decreases with gain 
of quark mass, but this phenomenon is not bound with mass. 
Fig.\ref{fig:excess} and existence of two-quark mesons with a $b$-quark 
are sufficient arguments about secondary role of quark mass  
in this respect.
This property of $c$-quark, more probably, is bound somehow with its electrical charge,
i.e. with structure of charges (1/3e + 1/3e) 
and of magnetic field inside a quark.
It is appropriate mention here that
the $u$-quark demonstrates  
a splittable structure of electric charge~\cite{my4, my5}.
In six-color six-quark configuration the gluon exchange of $c$-quark 
is much more because of additional interaction with other light quarks. 
At addition to this exchange there will an intensive background 
of gluon exchange between five light quarks.
Thus, it is possible that with the help of additional quark pairs 
the next properties will be reached:

1. An intensive gluon exchange;

2. The necessary degree of an antishielding around a color charge of $c$-quark; 

3. An asymptotic freedom of the movable $u$-quarks;

The result is the six-quark $D$-mesons
with masses 1864.5, 1869.3 and 1968.2 MeV. 

The reader, we might summarize aforesaid in other words, 
{\bf the shielding 
by the real quark-antiquark pair is the necessary condition 
for comfortable life of $c$-quarks in mesons (hadrons?)}.

Similar reasonings can be valid and for other mesons 
which contain $c$-quark(s), and also, perhaps, 
for their excited states (resonances).   
Still we not known the reason why $c$-quark should consist 
in six-color quark group. 
Perhaps, it is bound with features of its exchange of color gluons
and/or its acceptor ability.
Anyhow, it is bound with especial properties of $c$-quark only.   

\section {Conclusion}

\quad
Thus, the harmonic quarks perfectly work with first ground mesons 
of all flavors. It gives us the quark structures 
and exact additional energies of the next ground mesons:
\pion, \ip, \ak, \kaon, $D^0$, $D^\pm$, $D_s^\pm$, $\psi/J$, $B^0$ 
and $B^\pm$. All quarks of these mesons are a weak relativistic objects
as it have to be for ground states. 
Only three mesons with $b$-quarks ($B_s$, $B_c^\pm$ and $\Upsilon$(1S)) 
wait their turn.
However, this investigation shall more difficult because 
a $b$-quark is a donor (see fig.\ref{fig:excess}) 
and while we not know the quantity of energy 
which $b$-quark may give away.  

The QCD lattice simulation can be very efficient 
for an examination of new phenomenons and a development of 
harmonic quark model.

\end {document}